\documentclass[twocolumn,english,aps,prb,superscriptaddress]{revtex4}
\usepackage{amsmath}
\usepackage[T1]{fontenc}
\usepackage[latin9]{inputenc}
\usepackage{graphicx}
\usepackage{amssymb}
\usepackage{bm}
\usepackage{babel}

\input{epsf}

\begin{document}

\title{Optical transparency of graphene as determined by the fine-structure constant}
\author{Daniel E. Sheehy}
\affiliation{Department of Physics and Astronomy, Louisiana State University, Baton
Rouge, Louisiana 70803, USA}
\author{Jörg Schmalian}
\affiliation{Department of Physics and Astronomy and Ames Laboratory, Iowa State
University, Ames, Iowa 50011, USA }
\date{June 28, 2009}

\begin{abstract}
The observed $97.7\%$ optical transparency of graphene has been linked to
the value $1/137$ of the fine structure constant, by using results for
noninteracting Dirac fermions. The agreement in three significant figures
requires an explanation for the apparent unimportance of the Coulomb
interaction. Using arguments based on Ward identities, the leading
corrections to the optical conductivity due to the Coulomb interactions are
correctly computed (resolving a theoretical dispute) and shown to amount to
only $1$-$2\%$, corresponding to $0.03$-$0.04\%$ in the transparency.
\end{abstract}

\maketitle

The optical transparency of graphene is % fully
determined by its optical conductivity $\sigma\left(\omega\right) $ and $c $%
, the speed of light~\cite{Stauber08}: 
\begin{equation}
t\left( \omega \right) =\left( 1+2\pi \sigma\left(\omega\right) /c\right)
^{-2}.  \label{Eq:tee}
\end{equation}%
Recent experiments~\cite{Nair08} on suspended graphene found $t\left( \omega
\right) \simeq 0.977$, independent of $\omega$, in the visual regime ($450%
\mathrm{nm}<\lambda <750\mathrm{nm}$). This observation (see also Refs.~%
\onlinecite{Li08,Kuzmenko08,Mak08}) can be elegantly rationalized in terms
of non-interacting %massless
Dirac particles with optical conductivity~\cite{Ludwig} $\sigma
^{\left(0\right) } \left( k_{B}T\ll \omega \ll D\right) =\frac{\pi }{2}%
e^{2}/h$ . Here $D$ is the upper cut off energy for the linear dispersion,
of order of several electron volts and $T$ is the temperature. Assuming $%
\sigma =\sigma ^{\left( 0\right) }$ yields $t\left( \omega \right) \simeq
0.9774629(2)$, in excellent agreement with experiment. Thus, the optical
transparency of \textit{non-interacting\/} graphene $t\left( \omega \right)
=\left( 1+\pi \alpha _{\mathrm{QED}}/2\right) ^{-2}$ is solely determined by
the value of the fine structure constant of quantum electrodynamics: $\alpha
_{\mathrm{QED}}=e^{2}/\left( \hbar c\right) \simeq 1/137.035999(6)$. \
Despite the beauty of this reasoning, a natural question emerges: \textit{%
Why can one ignore the electron-electron Coulomb interaction?} After all the
Coulomb interaction in graphene is poorly screened and its strength is
governed by its own, effective fine structure constant $\alpha =e^{2}/\left(
\hbar v\right) \simeq 2.2$ that is significantly larger than $\alpha_{%
\mathrm{QED}} $ because of the smaller velocity~\cite{Novoselov05} $v\simeq
10^{6}\mathrm{m/s}$. The quantitative agreement 
%in three significant digits 
between experiment and a non-interacting theory clearly requires a
quantitative analysis of the size of interaction corrections to the optical
conductivity and transparency of graphene. %
%
% how
%accurately $\alpha_{\rm QED}$ is determined by the optical
%transparency of graphene.

%------------------------------
\begin{figure}[tbp]
\epsfxsize=9cm \vskip0.25cm \centerline{\epsfbox{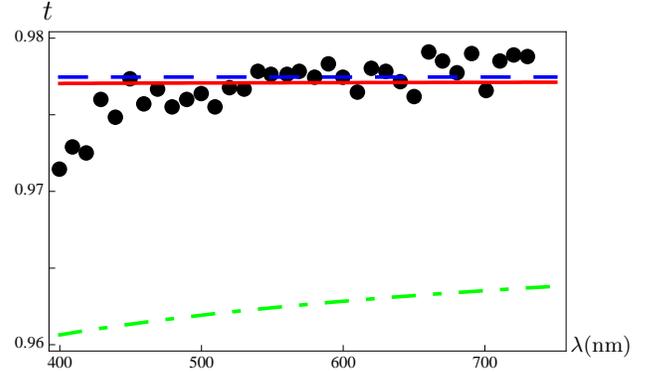}} \vskip-.55cm %
\vskip-.15cm
\caption{(Color Online) Optical transparency, Eq.~(\protect\ref{Eq:tee}), of
graphene, from Ref.~\onlinecite{Nair08} (points) along with theoretical
curves for the case of interacting graphene within the present theory, Eq.~(%
\protect\ref{eq:finalsigma}) with $\mathcal{C}_1 \simeq 0.01$ (solid red
line), according to the theory of Ref.~\onlinecite{Herbut08} ($\mathcal{C}_1
\simeq 0.51$; dot-dashed green line), and for noninteracting Dirac fermions
(dashed blue line).}
\label{opticalfig}\vskip-.45cm
\end{figure}
%------------------------------

In this Brief Report we determine the leading interaction corrections to the
optical transparency and demonstrate that they amount to only $0.037\% $ in
the visual regime. This surprisingly small correction is the consequence of
i) a perfect cancellation of the divergent (i.e. proportional to $\ln
D/|\omega|$) parts of Feynman diagrams that contribute to the conductivity
and ii) a near cancellation of the non-divergent contributions. While the
first result has been stated earlier by us~\cite{Sheehy07} as well as in
Ref.~\onlinecite{Herbut08}, the latter effect has been a subject of a
dispute~\cite{Herbut08,Mishchenko08,comment}. Below we resolve this dispute
and demonstrate that the leading perturbative correction to the conductivity
was correctly analyzed by Mishchenko in Ref.~\onlinecite{Mishchenko08}. We
show that perturbative corrections to the conductivity must be obtained by
guaranteeing that momentum cutoffs, used to regularize divergences, are
introduced in a fashion that respects Ward identities and thus guarantees
charge conservation.

The low energy Hamiltonian for electrons in graphene~\cite{CastroNeto} is
obtained by expanding to leading order in gradients near the nodes of the
tight-binding dispersion, %
% $\mathbf{k}_{1}=\frac{4\pi }{3a_{0}}\left( 3^{-1/2},0\right) $
%or $\mathbf{k}_{2}=\frac{2\pi }{3a_{0}}\left( 3^{-1/2},-1\right) $,
%
yielding the following nodal-fermion Hamiltonian: 
\begin{equation}
H\!=\!v \sum_{\mathbf{k},i}\psi_{i}^{\dagger }(\mathbf{k}) \mathbf{k}\cdot %
\bm{\sigma}\psi_{i}^{\phantom{\dagger}}(\mathbf{k})+ \frac{e^{2}}{2}\int
d^{2}rd^{2}r^{\prime }\frac{\rho (\mathbf{r}) \rho(\mathbf{r}^{\prime })}{%
\left\vert \mathbf{r}-\mathbf{r}^{\prime} \right\vert }. 
\end{equation}%
Here $i=1\cdots N$, with $N=4$ counting the two spin indices and two
independent nodes in the Brillouin zone, $\sigma_\mu$ are the Pauli
matrices, and we have set $\hbar= 1$. It is simple to show that the charge
density $\rho\left(\mathbf{r}\right) =\sum_{i=1}^{N}\psi_{i}^{\dagger }\left(%
\mathbf{r}\right) \psi_{i}^{\phantom{\dagger}} \left( \mathbf{r}\right) $
and current density $\mathbf{j\left( \mathbf{r}\right) =}v\sum_{i=1}^{N}\psi
_{i}^{\dagger }\left( \mathbf{r}\right) \bm{\sigma}\psi_{i}^{%
\phantom{\dagger}}\left( \mathbf{r}\right)$ are related by the continuity
equation: 
\begin{equation}
\frac{\partial \rho }{\partial t}+\mathbf{\nabla }\cdot \mathbf{j}=0.
\label{Eq:continuity}
\end{equation}%
The optical conductivity is related by the Kubo formula to the retarded
current-current correlation function: 
\begin{equation}
\sigma (\omega )=\frac{e^{2}}{2\omega }\mathrm{Im}\big[f_{xx}^{R}(\mathbf{q=0%
},\omega )+f_{yy}^{R}(\mathbf{q=0},\omega )\big].  \label{Kubo}
\end{equation}%
%
%
%
%where $\mu =x$ or $y$. 
Here, $f_{\mu \nu }^{R}(\mathbf{q},\omega )$ is the retarded correlation
function determined from the Matsubara function $f_{\mu \nu }(Q
)=\left\langle j_{\mu }\left( Q\right) j_{\nu }\left( -Q\right)
\right\rangle $ by analytically continuing $i\Omega \rightarrow \omega
+i0^{+}$. We use the convention $Q=\left( -i\Omega ,\mathbf{q}\right) $ and
correspondingly write $j_{0}\left( Q\right) =\rho \left( Q\right) $ for the
charge density.

The theory for the optical conductivity in graphene with electron-electron
Coulomb interaction was developed in Refs.~\onlinecite{Sheehy07,Herbut08}.
Using renormalization group (RG) arguments it holds that the effective
fine-structure constant of graphene, $\alpha $, becomes a running coupling
constant $\alpha \left( l\right) $ where $l$ is the flow variable of the RG
approach. In the case of graphene $\alpha \left( l\right) $ decreases
logarithmically as one lowers the typical energy scale~\cite%
{Gonzalez99,Ye98,gorbar,Son07,Sheehy07,Herbut08,Fritz08}. The optical
conductivity $\sigma \left( \omega ,T,\alpha \right) $ at frequency $\omega $%
, temperature $T$ and for the physical coupling constant $\alpha $ is
related to its value at a rescaled frequency $\omega _{R}\left( l\right)
=Z\left( l\right) ^{-1}\omega $, rescaled temperature $T_{R}\left( l\right)
=Z\left( l\right) ^{-1}T$ as well as the running coupling constant via 
\begin{equation}
\sigma \left( \omega ,T,\alpha \right) =\sigma \left( \omega _{R}\left(
l\right) ,T_{R}\left( l\right) ,\alpha \left( l\right) \right) .
\label{scaling}
\end{equation}%
The scaling factor up to one loop is given by $Z\left( l\right)
=e^{-l}\left( 1+\frac{\alpha }{4}l\right) $. Equation~(\ref{scaling})
implies that the conductivity is scale invariant under the RG flow. This
result is true to arbitrary order in perturbation theory as can be shown
following arguments by Gross~\cite{Gross75}. It is physically due to the
fact that the electron charge is conserved~\cite{Sheehy09}. The scaling
functions $\omega _{R}\left( l\right) $ and $T_{R}\left( l\right) $ grow
under renormalization while $\alpha \left( l\right) $ decreases~\cite%
{Sheehy07,Sheehy09}. Thus, in the relevant collisionless regime $\omega \gg T
$ is it sufficient to analyze the high frequency ($\omega _{R}\left(
l\right) \simeq D$), weak coupling limit where 
\begin{equation}
\sigma \left( D,0,\alpha \right) =\sigma ^{\left( 0\right) }\left[ 1+%
\mathcal{C}_{1}\alpha +\mathcal{C}_{2}\alpha ^{2}+\cdots \right] .
\end{equation}%
Here, the numerical coefficients $\mathcal{C}_{i}$ are determined by
performing an explicit perturbation theory. The scaling law Eq.~(\ref%
{scaling}) yields the conductivity as a function of frequency where we
replace $\alpha $ by the running coupling constant 
\begin{equation}
\alpha \rightarrow \alpha \left( \omega \right) =\alpha /\left( 1+\frac{%
\alpha }{4}\log \left( D/\omega \right) \right) ,  \label{Eq:alpharg}
\end{equation}%
here obtained to leading logarithmic accuracy\cite%
{Gonzalez99,Ye98,gorbar,Son07,Sheehy07,Herbut08,Fritz08}. \ The result is
that interactions only give rise to additive corrections to $\sigma ^{\left(
0\right) }$ that are of the form 
\begin{equation}
\sigma \left( \omega \right) =\sigma ^{\left( 0\right) }\left[ 1+\mathcal{C}%
_{1}\alpha \left( \omega \right) +\mathcal{C}_{2}\alpha ^{2}\left( \omega
\right) +\cdots \right] .  \label{corrections}
\end{equation}%
Note, this behavior is correct in the collisionless regime $\omega \gg k_{B}T
$. Qualitatively different behavior occurs in the opposite, hydrodynamic
regime~\cite{Fritz08} $\omega \ll k_{B}T$.

Since $\alpha \left( \omega \rightarrow 0\right) =0$, it follows from Eq.~(%
\ref{corrections}) that $\sigma \left( \omega \rightarrow 0\right)
\rightarrow \sigma ^{\left( 0\right) }$. However, $\alpha \left( \omega
\right) $ only vanishes as $4/\log \left( D/\omega \right) $ and corrections
could easily be significant in the visible part of the spectrum where $%
\omega $ and $D$ are comparable. The dominant correction is due to the $%
\mathcal{C}_{1}\alpha $ term and will be analyzed in this Brief Report. Combining
Eqs.~(\ref{Eq:alpharg}) and (\ref{corrections}), and neglecting the
higher-order terms, we have 
\begin{equation}
\sigma \left( \omega \right) =\sigma ^{\left( 0\right) }\left[ 1+\frac{%
\mathcal{C}_{1}\alpha }{1+\frac{1}{4}\alpha \log \left( D/\omega \right) }%
\right] .  \label{eq:finalsigma}
\end{equation}%
Calculations of $\mathcal{C}_{1}$ were presented in Refs.~%
\onlinecite{Herbut08} and \onlinecite{Mishchenko08}, however with different
results. While the authors of Ref.~\onlinecite{Herbut08} obtained $\ \mathcal{C%
}_{1}=\left( 25-6\pi \right) /12\simeq \allowbreak 0.513$, Mishchenko~\cite%
{Mishchenko08} obtained a significantly smaller value $\mathcal{C}_{1}=\left(
19-6\pi \right) /12\simeq \allowbreak 0.01\allowbreak 25\ $, which was
however disputed in Ref.~\onlinecite{comment}. Determining the correct value
of $\mathcal{C}_{1}$ is important for two reasons. First, there is no
obvious mistake in either Ref.~\onlinecite{Herbut08} or Ref.\onlinecite{Mishchenko08}. It is clearly important from a purely theoretical
point of view to settle this issue and set the criteria for correct
calculations of interaction effects in graphene. Second, as we discuss in
more detail below and illustrate in Fig.~\ref{opticalfig}, the coefficient
determined in Ref.~\onlinecite{Herbut08} is not consistent with experiment,
impling that qualitatively new phenomena or even higher order corrections
would have to be invoked to understand the observations of Ref.~%
\onlinecite{Nair08}.

%------------------------------
\begin{figure}[tbp]
\epsfxsize=8cm \vskip1.5cm \centerline{\epsfbox{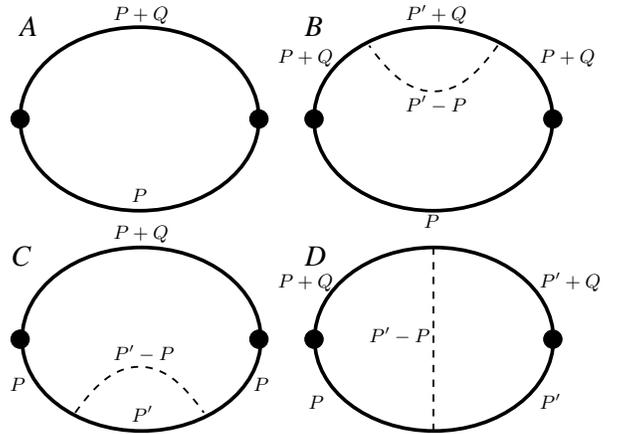}} \vskip-.35cm 
\vskip-.35cm
\caption{Feynman diagrams for the leading order contributions to $%
f_{\protect\mu\protect\nu}(Q)$ and $\protect\sigma(\protect\omega)$. Diagram
A is the $\mathcal{O}(\protect\alpha^0)$ contribution, while diagrams B, C,
D are $\mathcal{O}(\protect\alpha)$. Full lines represent fermions and
dashed lines represent the Coulomb interaction.}\vskip-.25cm
\label{Feynman}\vskip-.15cm
\end{figure}
%------------------------------

We now obtain $\sigma(\omega)$ by calculating the correlation function $%
f_{\mu\nu}(Q)$ which, as follows from Eq.~(\ref{Eq:continuity}), satisfies 
\begin{equation}
Q_{\mu }f_{\mu \nu }\left( Q\right) =0,  \label{cont}
\end{equation}
with the repeated index summed over $\mu = 0,x,y$.  When calculating $%
f_{\mu\nu}(Q)$, the leading contributions to which are shown in Fig.~\ref%
{Feynman}, we must ensure that Eq.~(\ref{cont}) is satisfied at each order
in $\alpha$.

The zeroth-order contribution, Fig.~\ref{Feynman}A, corresponds to the
current-current correlation function $f_{\mu \nu }^{\left( 0\right) }\left(
Q\right) $ of noninteracting Dirac particles and yields~\cite{Son07} 
\begin{equation}
f_{\mu \nu }^{\left( 0\right) }\left( Q\right) =\frac{N}{16\left\vert
Q\right\vert }\left( Q^{2}\delta _{\mu \nu }-Q_{\mu }Q_{\nu }\right) ,
\end{equation}%
which obeys Eq.~(\ref{cont}). Performing the analytic continuation
and inserting the result into the Kubo formula yields, after restoring
proper units, $\sigma ^{\left( 0\right) }=\frac{\pi e^2 }{2h}$ for $N=4$.
This leads to the $97.7\%$ optical transmission discussed above.

Next we analyze the three leading corrections to $\sigma ^{\left( 0\right) }$
as shown in Fig.~\ref{Feynman}B-D. The diagrams in Fig.~\ref{Feynman}B and C
yield the same contribution with interactions entering 
via self energy insertions with the leading self energy 
\begin{equation}
\Sigma (\mathbf{p})=- e^{2}\int_{P^{\prime }}V(\mathbf{p}-\mathbf{p}^{\prime
})G(P^{\prime }),  \label{self energy}
\end{equation}%
where $\int_{P^{\prime }}...=T\sum_{\omega ^{\prime }}\int_{p^{\prime }}%
\frac{d^{2}p^{\prime }}{(2\pi )^{2}}...$ and $V(\mathbf{p}) = \frac{2\pi}{|%
\mathbf{p|}}$ %
is the Fourier-transformed Coulomb interaction. The fermion propagator is
given by 
\begin{equation}
G\left( P\right) =\frac{-i\omega \sigma_{0}-v\mathbf{p}\cdot \mathbf{\sigma }%
}{\omega ^{2}+v^{2}p^{2}}.
\end{equation}%
The self energy, Eq.~(\ref{self energy}), diverges logarithmically and must
be regularized, for example by introducing an upper momentum cut off $%
\Lambda \simeq D/v$. %
%This divergence leads to the emergence of a running
%coupling constant once treated within an RG approach.
%
We will show that the discrepancy between previous calculations of $%
\sigma(\omega)$ can be traced to the fact that, in Eq.~(\ref{self energy}),
there are two obvious ways to introduce the ultraviolet (UV) cut-off $%
\Lambda $.

Thus, upon evaluating the frequency summation and momentum integrals, we
obtain 
\begin{equation}
\Sigma (\mathbf{p})=\frac{1}{4}\alpha v\mathbf{p}\cdot \bm{\sigma} \ln \left[
\frac{4\Lambda c}{p}\right],
\end{equation}
where the number $c$ depends on the cutoff procedure, with $c=\mathrm{e}%
^{-1/2}$ if we evaluate the momentum integral using the cutoff $\left\vert 
\mathbf{p}^{\prime }\right\vert <\Lambda$, i.e. by confining fermion states
to a circle near the node. (Note we always discard contributions that vanish
for $\Lambda/p\to \infty$). On the other hand, if we evaluate the momentum
integral by restricting $|\mathbf{p}-\mathbf{p}^{\prime }|<\Lambda$, i.e. by
having a finite Coulomb interaction at short distances (see also Ref.~%
\onlinecite{comment}), we find that $c=\mathrm{e}^{1/2}$. This corresponds
to replacing the Coulomb potential via $V(\mathbf{p})\rightarrow V_{\Lambda}(%
\mathbf{p})=2\pi \theta (\Lambda -|\mathbf{p}|)/|\mathbf{p}|$. We emphasize
that the log-divergent contribution to $\Sigma (\mathbf{p})$ is independent
of the regularization procedure, with the difference being in the subleading
contributions. 

As we will show, the two different values that have been determined for $%
\mathcal{C}_{1}$ in Refs.~\onlinecite{Herbut08} and
\onlinecite{Mishchenko08}, respectively, are directly related to the two
different values for $c$ in the self energy as it enters in the diagrams of
Fig.~\ref{Feynman}B and C. The diagram Fig.~\ref{Feynman}D is unaffected by
the regularization procedure. Which result for $c$, i.e. which
regularization procedure, is correct? The answer comes from the Ward
identity, which, as we show next, is only satisfied if we implement the
momentum cutoff by restricting the momenta in the Coulomb potential to $|%
\mathbf{p}-\mathbf{p}^{\prime }|<\Lambda $, implying that Mishchenko's
result~\cite{Mishchenko08} is correct.

To demonstrate that the proper cut off procedure is to restrict $|\mathbf{p}-%
\mathbf{p}^{\prime }|<\Lambda $ in the Coulomb potential, we analyze the
leading interaction corrections (Fig.~\ref{Feynman}B,C, and D), which we
call $f_{\mu \nu }^{\left( 1\right) }\left( Q\right) $. These satisfy: 
\begin{eqnarray}
Q_{\mu }f_{\mu \nu }^{(1)}(Q) &=&N\alpha \int_{P}\int_{P^{\prime
}}V_{\Lambda }(\mathbf{p}-\mathbf{p}^{\prime })  \label{Eq:bigward} \\
&&\times \mathrm{Tr}\left\{ G(P^{\prime }+Q)G(P+Q)G(P^{\prime }+Q)\sigma
_{\nu }\right.   \notag \\
&&\left. \hskip1cm-G(P^{\prime })\sigma _{\nu }G(P^{\prime })G(P)\right\} . 
\notag
\end{eqnarray}%
This result was obtained from the three diagrams Fig.~\ref{Feynman}B-D by
simply using the identity 
\begin{equation}
G(P)\!\left( i\Omega \sigma _{0}-\!\mathbf{q\cdot \sigma }\right)
G(P+Q)=G(P)-G(P+Q),  \label{Eq:identity}
\end{equation}%
and the cyclic property of the trace. At this point, the UV cutoff only
enters via $V_{\Lambda }(\mathbf{p})$, so that we can shift $P\rightarrow P-Q
$ and $P'\to P'-Q$ in the first term, again use the cyclic
property of the trace, and obtain 
\begin{equation}
Q_{\mu }f_{\mu \nu }^{(1)}(Q)=0,  \label{Eq:required}
\end{equation}%
the required result. Note that other regulation schemes for the UV behavior 
%(such as restricting the momentum arguments of the Green functions)
will \textit{not\/} necessarily work in this way. In particular, regulating
the momenta by restricting the Green-function momentum arguments amounts to
replacing $G(P)\rightarrow G(P)\Theta (\Lambda -|\mathbf{p}|)$; with such a
replacement, Eq.~(\ref{Eq:identity}) and thus Eq.~(\ref{Eq:bigward}) will
not be valid. We conclude, then, that in graphene momenta must be
regularized using $V_{\Lambda }(\mathbf{p})$.

The same conclusion can be arrived at by considering the leading corrections
to the current vertex 
\begin{equation}
\Lambda _{\mu }(P,Q)=-\alpha \int_{K}G(K)\sigma _{\mu }G(K+Q)V_{\Lambda }(%
\mathbf{k}-\mathbf{p}).
\end{equation}%
Again using Eq.~(\ref{Eq:identity}), we obtain 
\begin{eqnarray}
Q_{\mu }\Lambda _{\mu } &=&-\alpha \int_{K}\left( G(K+Q)-G\left( K\right)
\right) V_{\Lambda }(\mathbf{k}-\mathbf{p}),  \notag \\
&=&\Sigma (P+Q)-\Sigma (P),
\end{eqnarray}%
which is the correct Ward identity. Once again, alternate schemes for
cutting off the momentum integrals are not guaranteed to yield a proper Ward
identity of this form. 
%In particular, it is not clear how to implement a consistent Ward
%identity of this form by cutting off the arguments of the Green functions.
Our finding that a regularization in terms of a hard fermion cut-off
violates charge conservation is analogous to the observation in 
%quantum electrodynamics
QED that incorrect regularization schemes yield unphysical results such as a
photon mass~\cite{Mandl}.

Our final tasks are to evaluate the contributions to the current-current
correlation function and to determine the conductivity using the Kubo
formula Eq.~(\ref{Kubo}). \ We first consider the diagrams B and C, which
are identical. Recognizing the self-energy insertion, we have (with an
overall $2$ for the two diagrams): 
\begin{eqnarray}
\hspace{-.3cm}f_{\mu \nu }^{(1)}|_{BC}\!=\!-2N\!\int_{P}\!\mathrm{Tr}%
[G(P)\sigma _{\mu }G(P+Q)\sigma _{\nu }G(P)\Sigma (\mathbf{p})].
\end{eqnarray}
Evaluating the trace, and performing the frequency integral and analytical
continuation yields 
\begin{equation}
\mathrm{Im}f_{\mu \mu }^{(1)R}(\mathbf{q=0},\omega )|_{BC}=-\frac{N\alpha
\omega }{16}\ln \frac{8v\Lambda  ce^{-1/2}}{\omega },  \label{Eq:imfbc}
\end{equation}%
where the repeated $\mu $ index refers to the sum over the $xx$ and $yy$
components as in Eq.~(\ref{Kubo}). %
%Remember, the constant $c=e^{1/2}$ in the correct way to evaluate the UV
%cutoff, and $c=e^{1/2}$ in the incorrect scheme.
%
Analyzing the diagram D of Fig.~\ref{Feynman}, which can be written as 
\begin{eqnarray}
\hspace{-0.5cm}f_{\mu \nu }^{(1)}|_{D}=-N\!\int_{P}\!\mathrm{Tr}[G(P)\Lambda
_{\mu }(P,Q)G(P+Q)\sigma _{\nu }],
\end{eqnarray}
it turns out that the result does not depend on the details of the
regularization procedure and yields 
\begin{eqnarray}
\hspace{-0.25cm}\mathrm{Im}f_{\mu \mu }^{(1)R}(\mathbf{q=0},\omega )|_{D}=
\frac{N\alpha \omega }{16}\left( \ln \frac{8v\Lambda }{\omega }+\frac{19-6\pi 
}{6}\right) .\label{Eq:imfd}
\end{eqnarray}
By examining Eqs.~(\ref{Eq:imfbc}) and (\ref{Eq:imfd}), it is clear that the
dependence on the high energy scale $\Lambda $ vanishes, in agreement with
general scaling arguments~\cite{Sheehy07,Herbut08}. %
%Combining the various terms for the first order correction, we obtain 
%\begin{equation}
%\frac{\mathrm{Im}f_{\mu \mu }^{(1)R}(\omega )}{\omega }=\frac{N\alpha }{16}%
%\left( \frac{19-6\pi }{6}-\ln ce^{-1/2}\right) ,
%\end{equation}%
%
%
Plugging these results into Eq.~(\ref{Kubo}) yields Eq.~(\ref{eq:finalsigma}%
) with coefficient 
\begin{equation}
\mathcal{C}_{1}=\frac{19-6\pi }{12}-\frac{1}{2}\ln ce^{-1/2}.
\end{equation}%
We indeed see that the correct cut-off procedure, with $c=e^{1/2}$, yields $%
\mathcal{C}_{1}=\frac{19-6\pi }{12}$, whereas the other cutoff procedure,
corresponding to $c=e^{-1/2}$, yields\cite{Herbut08} $\mathcal{C}_{1}=\frac{%
25-6\pi }{12}$.

We have verified~\cite{Sheehy09} that the same result holds within alternate
regularization procedures that do not use sharp cutoffs, as long as the Ward
identity is satisfied. For example, replacing the Coulomb interaction 
$V(\mathbf{r})\rightarrow V_{\eta }(\mathbf{r})=\frac{e^2 r_0^{-\eta}}{r^{1-\eta}}$
with $r_0$ a length scale and $\eta >0$ (putting the physical system, at $d=2$, slightly below its own
upper critical dimension) regulates all integrals in a way consistent with
Eq.~(\ref{Eq:required}). We obtain (for $\eta\to 0$)
\begin{eqnarray}
\mathrm{Im}f_{\mu \mu }^{(1)R}|_{BC} &=&-\frac{N\alpha \omega }{16}\left( 
\frac{1}{\eta }+\ln \frac{4v }{\omega r_0}-\gamma_{E}\right) ,
\\
\mathrm{Im}f_{\mu \mu }^{(1)R}|_{D} &=&\frac{N\alpha \omega 
}{16}\left( \frac{1}{\eta }+\ln \frac{4v }{\omega r_0}-\gamma_{E}+\frac{%
19-6\pi }{6}\right)   ,
\notag
\end{eqnarray}%
with $\gamma_{E}$ the Euler constant.
Once again, while the separate contributions diverge with $\eta \rightarrow 0
$, their sum is convergent and yields the coefficient $\mathcal{C}_{1}=\frac{%
19-6\pi }{12}$.

As we have discussed, Eq.~(\ref{eq:finalsigma}) implies that the correction
to $\sigma^{\left(0\right)}$ is small at low photon energies $\omega\ll D$
regardless of the value of $\mathcal{C}_{1}$. However, at larger (i.e.,
optical~\cite{Nair08,Mak08}) frequencies, the second term may become
significant, depending on the value of the number $\mathcal{C}_{1}$. Nair 
\textit{et al.\/}~\cite{Nair08} find the conductivity to be $\sigma
/\sigma^{(0)}=(1.01\pm 0.04)$. If we take the value $\mathcal{C}_{1}\simeq
0.513$ from Ref.~\onlinecite{Herbut08}, however, Eq.~(\ref{corrections})
predicts a large frequency-dependent correction to the conductivity that is
not consistent with these error bars, giving, for example at photon
wavelength $\lambda =600nm$ (or $\hbar \omega =2.07$eV), $%
\sigma/\sigma^{(0)}\simeq 1.667$, assuming the bandwidth $D=7.24$eV. In
contrast, using $\mathcal{C}_{1}=\allowbreak 0.01\allowbreak 25$ of Ref.~%
\onlinecite{Mishchenko08}, yields for the same parameters, $%
\sigma/\sigma^{\left(0\right)}\simeq 1.016$, consistent with the error bars
of Nair \textit{et al.\/}~\cite{Nair08}. In Fig.~\ref{opticalfig} we show
the optical transparency that result from both values for $\mathcal{C}_{1}$
(using bandwidth $D = 7.24$eV), along with the free-Dirac fermion result as
function of wavelength $\lambda$ in comparison with experiment.

Given the smallness of interaction corrections, with $\sigma/\sigma^{(0)}%
\simeq 1.016$ for the correct value of $\mathcal{C}_{1}$ (a correction
comparable to corrections due to the true tight-binding band structure), it
is unlikely that optical measurements will reveal interaction effects.
Electron-electron interactions are much more visible in the enhanced
diamagnetic response~\cite{Sheehy07} or the hydrodynamic transport~\cite%
{Fritz08}.

In summary, we determined the leading corrections to the optical
conductivity and transparency of graphene and find that they are very small
and determined by the fine structure constant $\alpha _{\mathrm{QED}}$ up to
corrections of order $1$-$2\%$ in the conductivity and $0.03-0.04\%$ in the
transparency. Correctly regularizing the UV-divergent contributions required
using Ward identity arguments to resolve previous discrepancies in recent
literature (a controversy that persists~\cite{Golub}). Our work demonstrates
that there are no discrepancies in $\sigma(\omega)$ obtained by different
theoretical methods~\cite{Mishchenko08}, such as the Kubo formula, the
density polarization approach or kinetic approaches, if charge conservation
is guaranteed at all stages of the calculation. %
% While recent work has commented that such 
% discrepancies~\cite{Mishchenko08} arise from whether one uses a Kubo formula or
% polarization function approach to compute $\sigma (\omega )$, we find that,
% if the UV cutoff is handled correctly, these approaches agree (as they
% must). 
%
Our methods confirm and, more importantly, justify  the result first
obtained by Mishchenko~\cite{Mishchenko08} and  provide a general
prescription for calculating interaction corrections in graphene.

\textit{Acknowledgments\/} --- %---------
We gratefully acknowledge useful discussions with I. Vekhter, as well as the
Aspen Center for Physics where part of this work was carried out. This
research was supported by the Ames Laboratory, operated for the U.S.
Department of Energy by Iowa State University under Contract No.
DE-AC02-07CH11358, and by the Louisiana Board of Regents, under grant No.
LEQSF (2008-11)-RD-A-10.

\end{document}